# A Generation-based Text Steganography Method using SQL Queries

Youssef Bassil
LACSC – Lebanese Association for Computational Sciences
Registered under No. 957, 2011, Beirut, Lebanon

## ABSTRACT
Cryptography and Steganography are two techniques commonly used to secure and safely transmit digital data. Nevertheless, they do differ in important ways. In fact, cryptography scrambles data so that they become unreadable by eavesdroppers; while, steganography hides the very existence of data so that they can be transferred unnoticed. Basically, steganography is a technique for hiding data such as messages into another form of data such as images. Currently, many types of steganography are in use; however, there is yet no known steganography application for query languages such as SQL. This paper proposes a new steganography method for textual data. It encodes input text messages into SQL carriers made up of SELECT queries. In effect, the output SQL carrier is dynamically generated out of the input message using a dictionary of words implemented as a hash table and organized into 65 categories, each of which represents a particular character in the language. Generally speaking, every character in the message to hide is mapped to a random word from a corresponding category in the dictionary. Eventually, all input characters are transformed into output words which are then put together to form an SQL query. Experiments conducted, showed how the proposed method can operate on real examples proving the theory behind it. As future work, other types of SQL queries are to be researched including INSERT, DELETE, and UPDATE queries, making the SQL carrier quite puzzling for malicious third parties to recuperate the secret message that it encodes.

## 1. INTRODUCTION
Today, with the breakthrough of the Internet, securing data and private digital assets is becoming more and more a necessity. As a result, different techniqueswere conceived to provide some sense of data confidentiality and to ensure that the information exchanged over the Internet is never disclosed by unauthorized users. In fact, cryptography was the premier choice to individuals, companies, and governments to secure and protect their digital data. Although with cryptography, the content of the encrypted information is completely scrambled, the fact that there is information being stored, transmitted, or shared is not [1]. Consequently, an eavesdropper can easily capture the ciphered information and try to break itto uncover its real content. Therefore, there is undoubtedly a need for a novel tactic that hides the very existence of information so that no one, apart from the sender and receiver, would notice their presence.Steganography is the ideal solution to achieve this new level of data secrecy [2]. In essence, steganography is a technique for hiding data such as messages into another form of data such as image files [3]. Basically, one communicating party can use steganography to conceal a secret message called covered data into an image file called carrier file. The carrier file is then sent to the other communicating party who would decipher it and recover the secret data hidden inside. If by any means, the carrier file is intercepted by eavesdroppers, it would simply look like a regular innocent image file; and thus, they will let it pass through. Obviously, the foremost requirement of steganography is that the carrier file, which holds the secret data, must show no visual signsor artifacts so as to avoid arousing suspicions that somedata are being communicated secretly [4]. In practice, many types of steganography exist including steganography for images, audio files, video files, text files, HTML files, and Internet protocols [5]. However, there is yet no known steganography application for query languages such as SQL queries which may allow hiding secret data into structured queries, instead of images and audio files.

This paper proposes a new steganography method for textual data. It encodes input text messages into SQL carriers made up of SELECT queries. In fact, the output SQL carrier is dynamically generated out of the input message using a dictionary of words organized into 65 categories,each of which represents a particular character in the language. In short, the message to hide is first broken down into single characters each of which is mapped to a randomword from a corresponding category in the dictionary. For instance, letter "H" is mapped to a random word from "category H", digit "5" is mapped to a random word from "category 5", and so forth. Ultimately, all input characters are transformed into output words which are then put together to form an SQL query.The dictionary is an array-based hash table whose content can be defined by the communicating parties; thus, the SQL carrier, for the same input message, can vary depending on the dictionary being used, making it so robust against stego-attacks and so challenging to be reverse-engineered by third parties.

## 2. BASICS OF STEGANOGRAPHY
Steganography is the art and science of communicating in a way which hides the very existence of the communication [6]. In current practice, steganography is used to hide secret data such as text messagesinto a carrier files such as images while maintaining the size and quality of the carrier file.Formally, a steganography model can be defined as exhibiting two processes: The first one is the covering process which is denoted by F(A, D)=C where A is the original media file, D is the data to hide, and C is the carrier file housing the secret data D. The second one is the uncovering process which is denoted by $F^{-1}(C)=D$, where C is the carrier file and D is the recovered data [7]. Figure 1 depicts the basic steganography model.

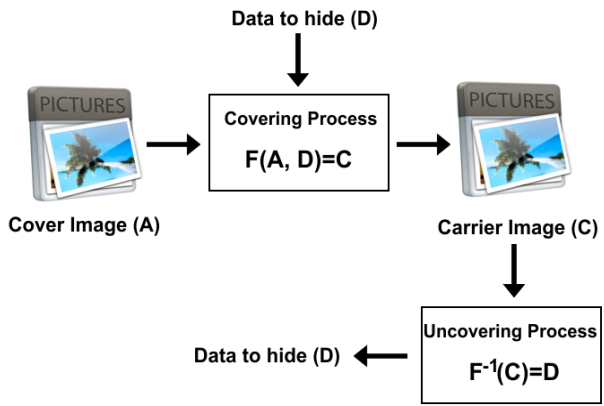

**Fig 1: Basic Steganography Model**

In the same context, steganography involves the following key elements [8]:

- The covert data (D) which represents the data to hide. D can be any type of data including plain text, document file, image file or any other bit stream file.
- The original media (A) into which dataDare to be hidden.
- The carrier file (C) which represents the resulting file which has the covert data D embedded into it.
- The stego function (F) and its inverse (F$^{-1}$), which represent the algorithms for hiding (covering) and recovering (uncovering) the covert data D from the carrier file C.

## 3. STATE-OF-THE-ART IN TEXT STEGANOGRAPHY

Hiding information in plain text can be done in many different ways. Some techniques consist of changing the outline of the carrier text such as adding whitespaces or altering the case of certain characters so as to represent secret text [9]. Others, consist of relating the characters to hide with the characters of the carrier text, creating a reference dictionary that maps words from the secret text with words from the carrier text [10]. This section sheds the light on the various techniques used in text steganography including hiding by selection [11], hiding in HTML [12], line and word shifting [13], hiding using whitespace [14], semantic-based hiding [15], and abbreviation-based hiding [16] techniques.

**Hiding by Selection**: The selection technique selects certain characters in the carrier text to convey the characters of the secret message such as selecting the first character of every word in the carrier text or the second character of every other word. Now, in order to recover the concealed secret message, all first characters of the words of the carrier text areextracted and concatenated together, producing the exact original message. A variation of this technique can be performed by selecting the first character from the first word, the second character from the second word, the third character from the third word, and so forth, until the characters of the message to hide are exhausted. NULL cipher [17] is in essence based on the selection technique as it constructs an unsuspicious plaintext having the secret message as part of its characters. For example, in World War I, the German embassy in United States sent a telegraph to Berlin stating "**P**resident's **E**mbargo **R**uling **S**hould **H**ave **I**mmediate **N**otice. **G**rave **S**ituation **A**ffecting **I**nternational **L**aw.**S**tatement **F**oreshadows **R**uin **O**f **M**any **N**eutrals. **Y**ellow **J**ournals **U**nifying **N**ational **E**xcitement **I**mmensely". It is by reading the first character of every word of this statement, that the secret message can be revealed as"Pershing Sails from NY June I" [18].The drawback of this method is that itrequires a huge volume of text to hide a small message of few words.

**HTML Documents**: Secret text can be easily concealed within HTML documents because HTML tags are case insensitive. For instance, the tags <a title="clients">, <a TITLE="clients">, and <a TitlE ="Center">, are all the same and have the same effect on the rendering of the document. Text steganography applied in HTML documents can be performed by changing the case of the letters that make up the HTML tags. In particular, the secret message is represented by the capital version of the tags' letters, or vice versa depending on the algorithm being used. As for the recovering process, all the capital letters from the HTML document have to be captured and concatenated together in order to produce the original covert message.

**Line and Word Shifting**: In this technique, text lines are shifted vertically and words are shifted horizontally by a fixed space of *n* inches. That way, the distance between lines and words would convey the hidden characters. For instance, letter A can be encoded as a 0.01 inch space between two text lines. Similarly, letter B can be encoded as a 0.02 inch space between two other text lines. In effect, this technique is more suitable for printed text than for digital text, since printed spaces can be physically measured unlike their digital counterparts.

**Hiding using Whitespace**: Its concept is very straightforward. A message to hide is first converted into a binary format. Then, every bit whose value is 1 is represented by an extra whitespace between a particular set of two words in the carrier text; whereas, every bit whose value is 0 leaves the original single whitespace between the next particular set of two words. For example, "the  boy went  to school today  " can be deciphered as "101001". In fact, two spaces exist between "the" and "boy", between "went" and "to", and between "today" and the end of the sentence. This results in a bit of value 1 in positions 0, 2, and 5 respectively. In contrast, only a single space exists between "boy" and "went", between "to" and "school", and between "school" and "today". This results in a bit of value 0 in positions 1, 3, and 4 respectively. Basically, the whitespace technique is very suspicious as a normal reader would right away notice the existence of some extra whitespaces in the text. Additionally, this method cannot encode too much information especially in small text.

**Semantic-Based Hiding:** Thesemantic-based text steganography technique uses synonyms of words to hide the secret information in the carrier text. For instance, the secret message "the boy went to school today" can be encoded using the semantic approach as "the child went to college today".

**Abbreviation-Based Hiding:** This technique uses a lexical dictionary containing words along with their abbreviations. These abbreviations are either labeled 0 or 1. While performing steganography, if a word in the carrier text is found in the dictionary, it is substituted by its abbreviation based on the current bit to hide. Different values of bits have different corresponding abbreviations. Some examples of these abbreviations can be "ASAP" for "As Soon As Possible", "CU" for "see you", "gr8" for "Great" etc. A variation of this method is the one proposed by [9], which consists of changing the spelling of words based on their American and British spellings. For example, "Favorite" is designated by 1 while "Favourite" is designated by 0, and "Center" is designated by 1 while "Centre" is designated by 0.

## 4. THE PROPOSED METHOD

This paper proposes a new text steganography method for hiding text messages into SQL carriers. In fact, it is a generation-based method that generates SQL queries [19] out of the secret message using a dictionary of words organized into 65 categories with no common words between these categories. In effect, these categories represent 65 different characters including the 26 letters of the English language, the 10 digits of the decimal system, and a set of 29 special characters. The proposed method uses the dictionary to decide on words to generate. Therefore, changing the content of the dictionary would result in a different SQL query for the same secret text message. This would make the algorithm more immune to stego-attacks and harder to be reverse-engineered by third parties.

Algorithmically, the message to hide is broken down into single characters each of which is mapped to a category of the dictionary. The obtained characters are then replaced by a random word from the corresponding category. Eventually, the characters of the message to hide are all transformed into words which are then put together to form an SQL query. Actually, the SQL query is the only data to be sent to the recipient as it represents the carrier text which encodes the secret message. The recipient on the other side has to revert the encoding operation so as to recover the secret text message out of the SQL carrier.

### 4.1 Properties of the Proposed Method

The proposed steganography method has several properties which are outlined below:

The first property is that the proposed method is a generation-based steganography which uses a dictionary of words to derive the output carrier, namely the SQL query. Consequently, changing the content of the dictionary would result in a different output SQL carrier for the same input secret message. This would make the method more resistant to stego-attacks and harder to be reverse-engineered by third parties.

The second property of the proposed method is the capacity of data that can be encoded using the SQL carrier. As a single character from the secret message is mapped into a single word from the dictionary, then the number of terms in the SQL carrier is equal to the number of characters in the secret data, in addition to other terms, mainly SQL keywords which are necessary to construct a correct SQL query such as "SELECT", "FROM", "AND", among other keywords. In total, an $n$ characters secret message would generate an $n$ terms SQL query + $k$ keywords.

The third property of the proposed method is that it is multilingual, in that, it can be custom-made to support languages other than the English language by just providing the appropriate dictionary for that language. As a result, the method can be used by a wide range of users regardless of their spoken language.

### 4.2 The Proposed Algorithm

The proposed algorithm comprises a list of steps to be undertaken in sequence in order to generate an SQL carrier out of a secret message. The steps are delineated below:

1. The message to hide denoted by $M=\{r_0,r_1,r_2,r_{m-1}\}$ where $r$ is a word in M and $m$ is the total number of words, is broken down into single characters yielding to $C=\{c_0,c_1,c_2,c_{n-1}\}$ where C is the sequence of characters that constitute the original message M, $c$ is a particular character, and $n$ is the total number of characters in C.

2. Every character $c_i$ is mapped to a random word $w_i$ from the dictionary whose category is equal to $c_i$. For instance, letter "H" is mapped to a random word from "category H", digit "5" is mapped to a random word from "category 5", and so forth. The words of the dictionary are stored in the form of $word=value$, for instance, $instructor="John"$.

   The mapping mechanism is in fact a generation process using a hash-based search algorithm [21] in which a hash value is computed for character $c_i$ using a hash function denoted by $h(c_i)$. Then the obtained value is used as an index to locate the category of the word $w_i$ in the dictionary. Once the category is found, $w_i$ is selected randomly out of this category. Mathematically, the hash function $h(c_i)$ is computed as $h(c_i) = ASCII(c_i) – 32$. Put differently, each character $c_i$ in C is mapped to a random word from the category whose index is equal to $h(c_i)=ASCII(c_i) – 32$. In fact, number 32 is used so as to obtain an index of 0 for the very first character, namely the SPACE whose ASCII value is 32.

3. The resulting generated words in step 2 are denoted by $W=\{c_0 \rightarrow w_0, c_1 \rightarrow w_1, c_2 \rightarrow w_2, c_{n-1} \rightarrow w_{n-1}\}$ where $c_i$ is a character in the original message mapped into $w_i$ which is a random word generated from the category whose index is equal to $h(c_i)=ASCII(c_i) – 32$. Finally, $n$ is the total number of the generated words in W which is also equal to the number of characters in the original message M.

4. The generated words in W are then split into two parts: A first part to represent terms of the SELECT clause, and a second part to represent terms of the WHERE clause of the SQL carrier. Formally, $W_{SELECT}=\{w_0...w_{(n-1)/2}\}$ and $W_{WHERE}=\{w_{((n-1)/2)+1}...w_{(n-1)}\}$.

5. A SELECT query with a WHERE clause is generated out of W. It has the following format: SELECT *select_clause* FROM *random_dummy_words* WHERE *where_clause*. The terms of the *select_clause* are comma separated, the *random_dummy_words* are just predefined dummy table names, and the *where_clause* is an ANDed terms clause. Besides, terms of the SELECT clause are selected from the dictionary without their values; while, terms of the WHERE clause are selected along with their original values, such as, SELECT *instructor* FROM *Table* WHERE *instructor="John"*. The word *instructor* is used without its value in the *select_clause*; whereas, it was used with its value in the *where_clause*.

6. For long SELECT queries, a postprocessor splits them into smaller queries based on a random index. This has no effect on recovering the secret message as long as the smaller queries are sent sequentially one after the other to the receiver.

Below is the pseudo-code for the proposed algorithm:

```
function hide(M) // M is the secret text to hide
{
    string W; // contains words to be mapped into SQL query
    char[] C ← ConvertToArrayChar(M)
    for (i←0 to i<length(C))
    {
        // finds the corresponding category index of character
        C[i]
```

```
        categoryIndex ← hash (dictionary, C[i]) // it returns
        ASCII(C[i])-32

        // choose a random word from this category
        W[i] ← dictionary[categoryIndex][randonIndex]
    }

    stringselect_qry ← "SELECT "
    for (i←0 to i<=(length(W))-1/2)
    {
        // choosing the right word along with its value
        select_qry ← select_qry + W[i] + "," ;
    }

    RemoveTrailingComma(select_qry);
    select_qry ← select_qry + " FROM " +
    tableNames[randomIndex] + " WHERE "

    for (i←((length(W))-1/2)+1 to i<= n-1)
    {
        // choosing the right word without its value
        select_qry ← select_qry + W[i].Split('=') + " AND " ;
    }
}
```

## 4.3 The Dictionary as a Hash Table

The dictionary is implemented as a hash table using two-dimensional array with rows corresponding to the categories of words and columns corresponding to the words in a particular category [20]. It comprises 65 categories with no common words between them. They represent 65 different characters including the 26 letters of the English language, the 10 digits of the decimal system, and a set of 29 special characters including @, #, $, ?, (, :, ., <, >, +, *, SPACE, etc. The number of words in each category (i.e. number of columns) is not predefined by the algorithm. It can range from one word to millions of words depending on user preference to choose the size of the dictionary. The more their numbers are, the harder is to recover the secret message from the SQL carrier. Table 1showshow the hash function can transform an input character $c$ into an output index value. This transformation is calculated as $h(c)=ASCII(c)-32$. It produces a hash value called index which points to the exact location of the category of character $c$.

**Table 1: Hash Function and the Characters of the Dictionary**

| c | Hash Function $h(C)=ASCII(C)-32$ | Hash Value Index |
|---|---|---|
| SPACE | → | 0 |
| ! | → | 1 |
| " | → | 2 |
| … | … | … |
| 0 | → | 16 |
| 1 | → | 17 |
| 2 | → | 18 |
| … | … | … |
| A | → | 33 |
| B | → | 34 |
| C | → | 35 |
| … | … | … |
| ` | → | 64 |

Table 2shows the dictionary as a two-dimensional array used as a hash table with rows representing the different 65 categories and columns representing the various words of the $i^{th}$ row.

**Table 2. Dictionary as HashTable**

| Categories Determined By hash function | Words Selected Randomly | | | |
|---|---|---|---|---|
| | [0] | [1] | [2] | [n-1] |
| [0] | car= "BMW" | id= 123 | age= 90 | … |
| [1] | course= "Security" | instructor= "John" | SSN= 12133343 | … |
| [2] | phone= 0122322 | firstname= "Steve" | lastname= "Doe" | … |
| … | … | … | … | … |
| [16] | orderID= 100 | customer= "Mike" | quantity= 90 | … |
| [17] | major= "CSI" | job= "engineer" | birthdate= "1990" | … |
| [18] | price= 55 | date= "1/1/2000" | ISBN= "2234400543" | … |
| … | … | … | … | … |
| [33] | type= "admin" | title= "Mr" | city= "Las Vegas" | … |
| [34] | state= "Georgia" | status= "done" | level= 10 | … |
| [35] | country= "UK" | file= "ak00.dat" | votes= 120 | … |
| … | … | … | … | … |
| [64] | expiry= "3/3/2001" | currency= "USD" | flag= "true" | … |

## 5. EXPERIMENTATION & RESULTS

An illustrated example is considered in this section which demonstrates how a secret text message can generate an SQL carrier using the proposed steganography method. Assuming that the secret message is M="KILL BOB", and that the dictionary is that of Table 2, the following results are obtained.

1. Message M is broken down into single characters such as C={$c_0,c_1,c_2,c_{n-1}$}={K, I, L, L, SPACE, B, O, B}

2. Each character $c_i$ in C is mapped to a random word from the category whose index is equal to $h(c_i)=ASCII(c_i)-32$. The obtained results are delineated below:

    h(K)=75-32=43 → index 43
    h(I)=73-32=41 → index 41
    h(L)=76-32=44 → index 44
    h(SPACE)=32-32=0 → index 0
    h(B)=66-32=34 → index 34
    h(O)=79-32=47 → index 47

    Choosing random words from each of the above calculated category indices leads to the following generated set of words:

    W={$c_0$→$w_0$, $c_1$→$w_1$, $c_2$→$w_2$, $c_{n-1}$→$w_{n-1}$}={ K→salary=2000, I→pages=512, L→bind="hard", L→discount=5, SPACE→status="done", B→id=123, O→state="Georgia", B→age=90}.

3. The generated words in W are then split into two parts: $W_{SELECT}$={ $w_0...w_{(n-1)/2}$} and $W_{WHERE}$={ $w_{((n-1)/2)+1}...w_{(n-1)}$}, with $W_{SELECT}$ having its terms' values discarded. Then, a SELECT query with a WHERE clause is generated out of W. This query is on the form of SELECT *select_clause* FROM *random_dummy_words* WHERE *where_clause*.

The obtained final queryisthe following: *SELECT salary, pages, bind, discount FROM Books, Items WHERE status="done" AND id=123 AND state="Georgia" AND age=90*

The values of terms "salary", "pages", "bind", and "discount" were removed because they are part of the SELECT clause.

4. The above generated SQL query is the actual carrier text to be sent to the intended receiver. It actually encodes the secret message.

To recover the secret message out of the SQL carrier, the receiver must be using the same algorithm and the same dictionary as the ones used by the sender. In fact, the process is so straightforward;terms of the *select_clause* and the *where_clause* are mapped back to their original categories in the dictionary, revealing consequently the original characters of the secret message. Back to the previous example, following is the back track to recover the original secret message "KILL BOB".That is:

W={$w_0 \to c_0$, $w_1 \to c_1$, $w_2 \to c_2$, $w_{n-1} \to c_{n-1}$}
W={ salary$\to$K, pages$\to$I, bind$\to$L, discount$\to$L, status="done" $\to$SPACE, id=123$\to$B, state="Georgia" $\to$O, age=90$\to$B}
C={K, I, L, L, SPACE, B, O, B}
{KILL BOB}=M

## 6. CONCLUSIONS & FUTURE WORK

This paper proposed a new generation-based steganography method for textual data. In essence, it generates an output SQL query as carrier text out of an input secret text message. The terms and parameters of the SQL query are selected from a dictionary of words organized into 65 categories representing English letters, decimal digits, and special characters. When experimented, the proposed method proved that it is workable andfeasible to be implemented. As for the characteristics, the proposed method is tolerable against various stego-attacks as it is based on a dynamic dictionary whose content can be altered by the communicating parties. Moreover, the method is multilingual as it supports hiding messages written in different languages not only English.

As future work, other types of SQL queries are to be investigated such as INSERT, DELETE, and UPDATE queries. That way, the SQL carrier would be a mix of multiple types of queries instead of one single type, making it so confusing for unauthorized third parties to break it and recover the secret message that it encodes.

## 7. ACKNOWLEDGMENTS
This research was funded by the Lebanese Association for Computational Sciences (LACSC), Beirut, Lebanon, under the "Stealthy Steganography Research Project – SSRP2012".